\documentclass[12pt,epsf,qsymbols]{article}
\pdfoutput=1
\usepackage[utf8]{inputenc}
\usepackage[normalem]{ulem}
\usepackage[english]{babel}
\usepackage{tabularx}
\usepackage{array}
\usepackage{graphics}
\usepackage[pdftex]{graphicx}
\usepackage{psfrag}
\usepackage{epsfig}
\usepackage{subfig}
\usepackage{amsmath}
\usepackage{mathtools}
\usepackage{amssymb}
\usepackage{bbold}
\usepackage{ulem}
\usepackage{setspace}
\usepackage{rotating}
\usepackage{colortbl}
\usepackage{longtable}
\usepackage{slashed}
\usepackage{braket}
\usepackage{lineno}
\makeatletter
\usepackage{textcomp}
\usepackage[usenames,dvipsnames]{xcolor}
\usepackage{relsize}
\usepackage{cite}

\usepackage{float}

\usepackage{bbm}
\usepackage{bm}
\usepackage{enumerate}
\usepackage{amsmath}
\usepackage{verbatim}

\usepackage{multirow}
\usepackage{arydshln}
\usepackage{slashed}

\usepackage{hyperref}
\hypersetup{colorlinks,citecolor=nicegreen,linkcolor=niceblue}
\hypersetup{colorlinks=true}

\usepackage{pifont}

\allowdisplaybreaks

\setlongtables

\newcommand{\bea}{\begin{eqnarray}}
\newcommand{\eea}{\end{eqnarray}}
\newcommand{\beq}{\begin{equation}}
\newcommand{\eeq}{\end{equation}}
\newcommand{\ec}{\end{center}}
\newcommand{\bc}{\begin{center}}

\newcommand{\pdir}{p\kern -5.2pt\raise 0.2ex\hbox {/}}

\newcommand{\vdir}{v\kern -5.75pt\raise 0.15ex\hbox {/}}
\newcommand{\kdir}{k\kern -5.75pt\raise 0.15ex\hbox {/}}
\newcommand{\epsdir}{\epsilon\kern -5.0pt\raise 0.15ex\hbox {/}}
\newcommand{\bvdir}{\bar{v}\kern -5.75pt\raise 0.15ex\hbox {/}}
\newcommand{\Ddir}{D\kern -7.75pt\raise 0.20ex\hbox {/}}
\newcommand{\Adir}{A\kern -7.75pt\raise 0.20ex\hbox {/}}
\newcommand{\ldir}{l\kern -5.0pt\raise 0.2ex\hbox{/}}
\newcommand{\varepsdir}{\varepsilon\kern -5.5pt\raise 0.15ex\hbox{/}}

\makeatother

\definecolor{niceblue}{rgb}{0.15,0.15,0.6}
\definecolor{nicegreen}{rgb}{0.1,0.5,0.1}
\definecolor{Red}{rgb}{1.,0.,0.}

\definecolor{Green}{rgb}{0.2,.7,0.2}

\begin{document}
\unitlength = 1mm

\thispagestyle{empty} 
\begin{center}
\vskip 3.4cm\par
{\par\centering \textbf{\LARGE  
\Large \bf Muon $g-2$ and scalar leptoquark mixing}}
\vskip 1.2cm\par
{\scalebox{.85}{\par\centering \large  
\sc Ilja Dor\v{s}ner$^a$, Svjetlana Fajfer$^b$ and Olcyr Sumensari$^{c}$}
{\par\centering \vskip 0.7 cm\par}
{\sl 
$^a$~{University of Split, Faculty of Electrical Engineering, Mechanical Engineering and Naval Architecture in Split (FESB), Ru\dj era Bo\v{s}kovi\'{c}a 32, 21000 Split, Croatia}}\\
{\par\centering \vskip 0.25 cm\par}
{\sl 
$^b$~{Department of Physics, University of Ljubljana, Jadranska 19, 1000 Ljubljana, Slovenia\\
Jo\v{z}ef Stefan Institute, Jamova 39, P. O. Box 3000, 1001 Ljubljana, Slovenia}}\\
{\par\centering \vskip 0.25 cm\par}
{\sl 
$^c$~Istituto Nazionale Fisica Nucleare, Sezione di Padova, I-35131 Padova, Italy\\
Dipartamento di Fisica e Astronomia ``G.~Galilei", Università di Padova, Italy}\\

{\vskip 1.65cm\par}}
\end{center}

\vskip 0.85cm
\begin{abstract}
The observed muon anomalous magnetic moment deviates from the Standard Model predictions. There are two scalar leptoquarks with simultaneous couplings to the quark-muon pairs of both chiralities that can singly explain this discrepancy. We discuss an alternative mechanism that calls for the mixing of two scalar leptoquarks of the same electric charge through the interaction with the Higgs field, where the two leptoquarks separately couple to the quark-muon pairs of opposite chirality structures. Three scenarios that satisfy this requirement are $S_1\,\&\, {S}_3$, $\widetilde{S}_1\,\&\, S_3$, and $\widetilde{R}_2\,\&\, R_2$, where the first scenario is realised with the up-type quarks running in the loops while the other two scenarios proceed through the down-type quark loops. We introduce only two non-zero Yukawa couplings to the relevant quarks and a muon, at the time, to study ability of these three scenarios to explain $(g-2)_\mu$ and be in accord with available experimental constraints. We find that the $S_1\,\&\, {S}_3$ scenario with the top-quark loops is consistent with all existing measurements. The $\widetilde{S}_1\,\&\, S_3$ and $\widetilde{R}_2\,\&\, R_2$ scenarios can accommodate the observed discrepancy through the bottom-quark loops but exhibit significant tension with the existing data on the high-$p_T$ dilepton-tails at LHC for the required values of Yukawa couplings and leptoquark masses. 
\end{abstract}
\newpage
\setcounter{page}{1}
\setcounter{footnote}{0}
\setcounter{equation}{0}
\noindent

\renewcommand{\thefootnote}{\arabic{footnote}}

\setcounter{footnote}{0}

\tableofcontents

\newpage

\section{Introduction}
\label{Sec:intro}

Leptoquarks (LQs) are hypothetical particles of either scalar or vector nature that couple at the tree-level to quark-lepton pairs~\cite{Buchmuller:1986zs,Dorsner:2016wpm}. The fact that they interact with the fields with well-defined baryon ($B$) and lepton ($L$) numbers makes them particularly appealing sources of New Physics (NP) regarding phenomena related to the $B$ and/or $L$ number violation. These phenomena include proton decay~\cite{Nath:2006ut} and generation of neutrino mass~\cite{Chua:1999si,Mahanta:1999xd} to name a few.

In this work we are interested in the $B$ and $L$ number conserving effects of scalar LQs with regard to the anomalous magnetic moment of muon. The main reason behind this study is the long-standing discrepancy between the measured value~\cite{Bennett:2006fi} and theoretical predictions of that observable~\cite{Jegerlehner:2009ry,Keshavarzi:2018mgv,Davier:2019can}. The experimental result ($a_\mu^\mathrm{exp}$) for the muon anomalous magnetic moment deviates from the Standard Model (SM) predictions ($a_\mu^\mathrm{th}$) roughly at the level of 4\,$\sigma$~\cite{Jegerlehner:2018gjd}. More precisely, the discrepancy that we want to address currently reads
\begin{equation}
\label{amu}
\delta a_\mu= a_\mu^\mathrm{exp} - a_\mu^\mathrm{SM}= (2.7 \pm 0.8)\times 10^{-9}\,.
\end{equation}
Our interest is further amplified by the anticipated experimental result from the Muon g$-$2 Collaboration~\cite{Grange:2015fou} at Fermilab that might clarify the nature of the current disagreement.

\begin{figure}[b]
\centering
\includegraphics[width=0.75\textwidth]{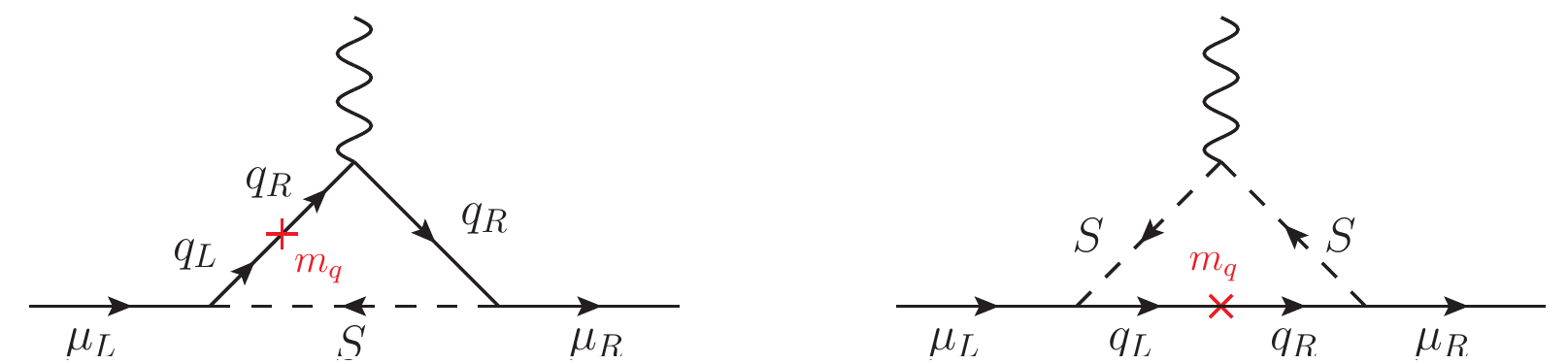}
\caption{The chirality-enhanced one-loop contributions to muon dipoles ($\propto m_q/m_\mu$) due to a presence of scalar $S$ that couples to both left- and right-chiral muons, where $S$ is either $R_2$ or $S_1$ and $q \in \lbrace u,c,t \rbrace$.}
\label{fig:diagram-lq-single}
\end{figure}

The influence of scalar LQs on $a_\mu=(g-2)_\mu/2$ is well-documented in the literature~\cite{Djouadi:1989md}. The only contribution that might be large enough to address the observed difference is of the one-loop nature and it requires a presence of a non-chiral LQ~\cite{Cheung:2001ip}. The non-chiral LQs are those that couple to both left- and right-chiral quarks of the same type and, due to Lorentz symmetry, to the charged leptons of opposite chiralities, where we assume that the fermion content of the NP scenario is purely the SM one. It turns our that the only scalar LQ multiplets that possess non-chiral couplings to muons are $R_2$ and $S_1$. We specify, for completeness, transformation properties of all scalar LQs under the SM gauge group $SU(3)_c \times SU(2)_L \times U(1)_Y$ and associated nomenclature in Table~\ref{tab:LQs}. We also show the generic one-loop Feynman diagrams induced by the non-chiral scalar LQs that yield contributions to $(g-2)_\mu$ in Fig.~\ref{fig:diagram-lq-single}. Clearly, the SM fermions that close the $(g-2)_\mu$ loop, in the non-chiral LQ case, are the up-type quarks while the leading NP contribution is proportional to the mass of the quark in the loop and inversely proportional to the square of the mass of the propagating LQ.

\begin{table}[t]
\renewcommand{\arraystretch}{1.45}
\centering
\begin{tabular}{|c|c|c|c|}
\hline
Symbol &  $(SU(3)_c,SU(2)_L,U(1)_Y)$  &  Interactions &  $F= 3B+L$ \\
\hline
\hline 
$S_3$ &	$(\overline{\mathbf{3}},\mathbf{3},1/3)$ & $\overline{Q}^C L$ & $-2$\\
$R_2$ &	$(\mathbf{3},\mathbf{2},7/6)$ & $\overline{u}_R L$,~$\overline{Q}e_R$ & $0$\\
$\widetilde{R}_2$ &	$(\mathbf{3},\mathbf{2},1/6)$ & $\overline{d}_R L$ & $0$\\
$\widetilde{S}_1$ &	$(\overline{\mathbf{3}},\mathbf{1},4/3)$ & $\overline{d}^C_Re_R$ & $-2$\\
$S_1$ &	$(\overline{\mathbf{3}},\mathbf{1},1/3)$ & $\overline{Q}^C L$,~$\overline{u}_R^C e_R$ & $-2$\\	\hline
\end{tabular}
\caption{\label{tab:LQs} List of all scalar LQs, their SM quantum numbers, renormalizable interactions to the quark-lepton pairs, and associated fermion numbers. Interactions with right-handed neutrinos are not considered.}
\end{table}

In this work we investigate the viability of those scenarios where the one-loop contributions towards the anomalous magnetic moment of muon are induced through the mixing of two scalar LQs of the same electric charge via the SM Higgs field, where the LQs in question need to couple to the muons of opposite chiralities. We accordingly study the existing constraints on the parameter space of this particular mechanism due to electroweak precision measurements, relevant flavor observables, and the current LHC limits. 

The paper is organized as follows. In Sec.~\ref{sec:single-lq}, we describe single LQ contributions to $(g-2)_\mu$ to establish the notation. In Sec.~\ref{sec:lq-mixing}, we classify those pairs of scalar LQs that can mix via the SM Higgs field and subsequently generate chirality-enhanced contributions towards $(g-2)_\mu$. We find three possible LQ pairs -- $S_1\,\&\, {S}_3$, $\widetilde{S}_1\,\&\, S_3$, and $\widetilde{R}_2\,\&\, R_2$ -- that might yield large enough contributions towards $(g-2)_\mu$ through the mixing with the SM Higgs field. We proceed to discuss electroweak precision constraints on the LQ mixing and discuss relevant differences between the three scenarios, if any. We then confront, in Sec.~\ref{sec:gminus2}, the $S_1\,\&\, {S}_3$ scenario with $(g-2)_\mu$ and various phenomenological constraints to investigate its viability. The ability of the $\widetilde{S}_1\,\&\, S_3$ and $\widetilde{R}_2\,\&\, R_2$ scenarios to address $(g-2)_\mu$ is briefly discussed in Secs.~\ref{sec:others_2} and~\ref{sec:others_1}, respectively. We summarize our findings in Sec.~\ref{sec:conclusion}.

\section{Single LQ contributions to $(g-2)_\mu$}
\label{sec:single-lq}

The most general formulae for the interactions of the generic scalar LQ $S$ of the definite fermion number $F=3B+L$ with the quark-charged lepton ($q$-$\ell$) pairs, in the mass eigenstate basis, are~\cite{Djouadi:1989md}
\begin{align}
\begin{split}
{\cal L}^{F=0} &= \overline{q}_i\,( l^{ij} P_R + r^{ij} P_L )\, \ell_j\, S +\mathrm{h.c.}\\[0.35em]
 {\cal L}^{|F|=2} &= \overline{q}_i^C\, ( l^{ij} P_L + r^{ij} P_R)\, \ell_j\, S +\mathrm{h.c.}\,,
\end{split}
\end{align}
where $i$ represents generation index for quarks, $j$ is generation index for charged leptons, $P_{L,R}$ are projection operators, and $l^{ij}$ and $r^{ij}$ are Yukawa coupling strengths. The LQ contributions to the muon anomalous moment can then be written in the compact form \cite{Dorsner:2016wpm}
\begin{align}
\begin{split}
\delta a_\mu = - \frac{N_c m_\mu}{8 \pi^2 m_S^2} \sum_q &\Big{[} m_\mu (|l^{q \mu}|^2 + |r^{q \mu}|^2 ) \,\mathcal{F}_{Q_S}(x_q)+ m_q\, \mathrm{Re} (r^{q \mu*} \,l^{q \mu})\, \mathcal{G}_{Q_S}(x_q) \Big{]}\,,
\end{split}
\label{eq:amu}
\end{align}
where $m_\mu$ is muon mass, $m_S$ is the LQ mass, $m_q$ is the quark mass, $x_q=m_q^2/m_S^2$, $N_c=3$ is the number of colors, and
\begin{align}
\label{eq:F-Q}
\mathcal{F}_{Q_S}(x) &= Q_S\, f_S(x)-f_F(x)\,,\\[0.3em]
\label{eq:G-Q}
\mathcal{G}_{Q_S}(x) &= Q_S\, g_S(x)-g_F(x)\,.
\end{align}
We denote with $Q_S$ the electric charge of scalar $S$ and define the loop functions
\begin{align}
\label{eq:fS}
f_S(x) &= \frac{x+1}{4 (1-x)^2} + \frac{ x \log x}{2(1-x)^3}\,,\qquad\quad g_S(x) = \frac{1}{x-1} - \frac{\log x}{(x-1)^2}\,,\\[0.35em]
\label{eq:fF}
f_F(x) &= \frac{x^2-5x-2}{12(x-1)^3} + \frac{ x \log x}{2(x-1)^4}\,,\qquad g_F(x) = \frac{x-3}{2(x-1)^2} + \frac{\log x}{(x-1)^3}\,.
\end{align}
It is clear from Eq.~\eqref{eq:amu} that the scalar LQs that have only left- or right-chiral couplings generate contributions to $a_\mu$ that are of a definite sign and that are suppressed by the lepton mass. One thus minimally needs either one non-chiral LQ or two scalar LQs of the same electric charge that can mix through the Higgs field and that can couple to muons of opposite chiralities to generate phenomenologically viable shift in $a_\mu$. Again, according to the content of Table~\ref{tab:LQs}, the only LQs that can potentially generate Feynman diagrams shown in Fig.~\ref{fig:diagram-lq-single} are $R_2$ and $S_1$ with $q \in \lbrace u,c,t \rbrace$. 
 
There exists a number of dedicated studies of a single LQ resolution of the muon anomalous magnetic moment discrepancy~\cite{Queiroz:2014zfa,Biggio:2014ela,Bauer:2015knc,ColuccioLeskow:2016dox,Kowalska:2018ulj,Mandal:2019gff}. By replacing the respective values of $Q_S$ in Eqs.~\eqref{eq:fS} and \eqref{eq:fF} and by considering the top quark couplings, it is straightforward to see that the single LQ scenarios can explain the discrepancy in Eq.~\eqref{amu} via loops for perturbative Yukawas, i.e., $|l^{t\mu}|,|r^{t\mu}|<\sqrt{4\pi}$, and LQ masses as large as $\approx 100\,\mathrm{TeV}$. Conversely, for a fixed mass $m_S=1$\,TeV, the needed couplings are $\sqrt{|l^{t\mu}\, r^{t\mu}|} \approx 5\times 10^{-2}$. We, in this manuscript, turn our attention to an alternative explanation of $(g-2)_\mu$ within the LQ mixing scenario context that we define next. 

\section{LQ mixing scenarios}
\label{sec:lq-mixing}

\subsection{General classification}
\label{sec:ewpt}

We list in Table~\ref{tab:01} all possible pairs of scalar LQs that can mix through the SM Higgs field~\cite{Hirsch:1996qy} at the renormalizable level. We denote the SM Higgs field as $H=(\bm{1},\bm{2},1/2)$ and indicate the number of $H$ fields in the contraction in the second column of Table~\ref{tab:01}. Those LQ pairs that couple to the muons of opposite chiralities can generate the one-loop contribution towards $(g-2)_\mu$ that we are interested in whereas the pairs that couple to neutrinos can yield neutrino masses of Majorana nature~\cite{AristizabalSierra:2007nf,Dorsner:2017wwn,Cata:2019wbu} at the one-loop level. In both instances the LQ pairs need to couple to the quarks of the same type but opposite chiralities in order for the relevant loop(s) to be completed and, in the case of $(g-2)_\mu$, for associated contributions to be enhanced. We accordingly specify the quark type that is in the loop, where we use $u$ and $d$ to collectively denote the up-type and the down-type quarks, respectively. 
\begin{table}[H]
\renewcommand{\arraystretch}{1.45}
\centering
\begin{tabular}{|c|c|c|c|}
\hline
LQ pairs &  Mixing field(s)  &  $(g-2)_\mu$ &  $\nu$-mass \\
\hline
\hline 
$S_1\,\&\, S_3$ & $H\,H$ & $u$ & -- \\
$\widetilde{S}_1\,\&\, S_3$ & $H\,H$ & $d$ & -- \\
$\widetilde{R}_2\,\&\, R_2$ & $H\,H$ & $d$ & -- \\
$\widetilde{R}_2\,\&\, S_1$ & $H$ & -- & $d$ \\
$\widetilde{R}_2\,\&\, S_3$ & $H$ & -- & $d$ \\
\hline
\end{tabular}
\caption{\label{tab:01} Scalar LQ pairs that can, through the mixing with the SM Higgs field, generate either the one-loop contributions towards $(g-2)_\mu$ or neutrino mass. It is indicated whether the chirality-enhanced contributions are proportional to the up-type ($u$) or down-type ($d$) quark masses.}
\end{table}
There are, clearly, three possible LQ pairs that might generate large enough contributions towards $(g-2)_\mu$ through the mixing with the SM Higgs field. These combinations are $S_1\,\&\, S_3$, $\widetilde{S}_1\,\&\, S_3$, and $\widetilde{R}_2\,\&\, R_2$, where, in all three instances, at least one of the LQ multiplets is chiral in nature. Since at least one of the two LQ multiplets that mix carries non-trivial $SU(2)_L$ assignment, the LQ mixing mechanism induces the mass splitting between the states belonging to the same $SU(2)_L$ multiplet. This, in turn, can generate oblique corrections that might be constrained by the existing electroweak precision measurements. We find these constraints to be marginally relevant for those LQ masses and associated Yukawa couplings that are not in conflict with the results of the existing LHC analyses.

\subsection{Mixing formalism}
\label{ssec:formalism}

To describe the most prominent features of the LQ mixing we assume existence of two scalars $S^{(Q)}_a$ and $S^{(Q)}_b$, of the same electric charge $Q$, but from two different multiplets $S_a$ and $S_b$ that might have non-trivial weak isospins $I^{S_a}$ and $I^{S_b}$. We thus expect, on general grounds, to have $2 (I^{S_a}+I^{S_b}+1)$ mass eigenstates with or without mixing. The mass squared matrix for the mixed states reads
\begin{align}
\label{eq:mix_matrix}
\mathcal{M}^2 =  \begin{pmatrix}
m_{S_a}^2 &  \Omega \\
 \Omega & m_{S_b}^2
\end{pmatrix}\,,
\end{align}
where $m_{S_a}$ and $m_{S_b}$ denote the common masses of all $S_a$ and $S_b$ components prior to the mixing and $\Omega$ stands for the mixing term arising from the interactions of $S_a$ and $S_b$ with the Higgs boson that we discuss later on within concrete NP scenarios. The matrix in Eq.~\eqref{eq:mix_matrix} can be brought into diagonal form with a simple field redefinition
\begin{align}
\label{eq:mix}
 \begin{pmatrix}
 S^{(Q)}_- \\
 S^{(Q)}_+
 \end{pmatrix}
 = \begin{pmatrix} \cos \theta & \sin \theta \\
 -\sin\theta & \cos \theta\end{pmatrix}
  \begin{pmatrix}
 S^{(Q)}_a \\
 S^{(Q)}_b
 \end{pmatrix}\,,
\end{align}
where $S^{(Q)}_{\pm}$ are the mass eigenstates and the mixing angle $\theta\in [-\pi/4,\pi/4]$ is related to $\Omega$ via the relation
\begin{equation}
\label{eq:theta}
\tan 2\theta = \dfrac{2\,\Omega}{m_{S_a}^2-m_{S_b}^2}.
\end{equation}
The LQs mix maximally, according to Eq.~\eqref{eq:theta}, for $m_{S_a} = m_{S_b}$ with $|\theta| = \pi/4$. We also note that $m^2_{S_a},m^2_{S_b} > |\Omega|$ due to the actual origin of $\Omega$ in our set-up and the current limits from LHC on the LQ masses as we explicitly show later on. 

The physical masses squared for the mixed states are
\begin{equation}
\label{eq:mSpm}
m_{S^{(Q)}_\pm}^2=\dfrac{m_{S_a}^2+m_{S_b}^2}{2} \pm \dfrac{1}{2}\sqrt{(m_{S_a}^2-m_{S_b}^2)^2+4 \,\Omega^2}\,.
\end{equation}
where, for simplicity, we assume that $m_{S_b} \geq m_{S_a}$.~\footnote{The case $m_{S_a}\geq m_{S_b}$ can be obtained by the replacements $\theta\to -\theta$ and $S^+ \leftrightarrow S^-$ in Eqs.~\eqref{eq:mix} and \eqref{eq:theta}. In this paper, we assume $m_{S_b}\geq m_{S_a}$, unless stated otherwise.} There are additional $2 I^{S_a}$ ($2 I^{S_b}$) mass eigenstates in multiplet $S_a$ ($S_b$) with masses $m_{S_a}$ ($m_{S_b}$) besides the two mass eigenstates $S^{(Q)}_\pm$. 

The preceding expressions can now be straightforwardly applied to the three LQ combinations listed in Table~\ref{tab:LQs}. As noted before, the LQ mixing mechanism induces mass splitting between the states that belong to the same $SU(2)_L$ multiplet. We accordingly discuss associated constraint that arises from the electroweak precision data and, most particularly, the oblique parameter $T$, which is sensitive to the custodial symmetry breaking effects~\cite{Peskin:1991sw,Marciano:1990dp,Altarelli:1990zd,Kennedy:1991sn}.

\begin{itemize}
	\item[•]\underline{$S_1\,\&\, {S}_3$}: The interactions of $S_3=S_a$ and $S_1=S_b$ with the Higgs boson $H$ read
\begin{align}
\label{eq:mix-S1-S3}
\mathcal{L}^{S_1\,\&\, S_3}_{\mathrm{mix}} &= \xi\, H^\dagger (\vec{\tau}\cdot\vec{S_3})H S_1^\ast+\mathrm{h.c.}\,,
\end{align}
where $\xi$ is a dimensionless coupling that, after electroweak symmetry breaking, induces a mixing between the $Q=1/3$ states $S_3^{(1/3)}$ and $S_1\equiv S_1^{(1/3)}$. The mass eigenstates of the $Q=1/3$ fields are then described by Eq.~\eqref{eq:mix} for $S^{(Q)}_a = S_3^{(1/3)}$, $S^{(Q)}_b = S_1^{(1/3)}$, and $\Omega=-\xi \, v^2/2$. Since the weak isospins of LQ multiplets $S_1$ and $S_3$ are $I^{S_1}=0$ and $I^{S_3}=1$, respectively, this scenario has four mass eigenstates with masses $m_{S^{(1/3)}_\pm}$ and $m_{S_3}\equiv m_{S^{(4/3)}_3}=m_{S^{(-2/3)}_3}$.

The only relevant constraint on $\xi$ arises from the electroweak precision data and, most particularly, the oblique parameter $T$. By assuming $m_{S_1}\geq m_{S_3}$, we find that the modification of the $T$-parameter, i.e., $\Delta T = T-T^{\mathrm{SM}}$, is 
\begin{equation}
\Delta T_{S_1\,\&\, {S}_3} = \dfrac{N_c}{4\pi s_W^2}\dfrac{1}{m_W^2}\Bigg{[}\cos^2\theta\, F(m_{S_3},m_{S^{(1/3)}_-})+ \sin^2\theta\, F(m_{S_3},m_{S^{(1/3)}_+})\Bigg{]}\,,
\label{eq:T_1}
\end{equation}
with
\begin{equation}
F(m_1,m_2)= m_1^2+m_2^2 - \dfrac{2 m_1^2 m_2^2}{m_1^2-m_2^2}\log \left(\dfrac{m_1^2}{m_2^2}\right)\,,
\end{equation}
where we introduce the shorthand notation $\sin \theta_W = s_W$, with $\theta_W$ being the Weinberg angle. The function $F(m_1,m_2)$ is symmetric under $m_1 \leftrightarrow m_2$ and it satisfies $F(m,m)=0$. If we take $m_{S_1} \gg m_{S_3} > |\Omega|$, we obtain the decoupling limit of Eq.~\eqref{eq:T_1} that reads
\begin{equation}
\Delta T_{S_1\,\&\, S_3}= \dfrac{N_c}{4\pi s_W^2}\dfrac{1}{m_W^2} \dfrac{\Omega^2}{m_{S_1}^2}+\mathcal{O}\left(\dfrac{\Omega^2}{m_{S_1}^4}\right)\,.
\end{equation}

We impose the results from the most recent electroweak fit, i.e., $\Delta T =0.05(12)$~\cite{Baak:2012kk}, to ensure a viability of those values of $\xi$ that we use in our numerical analysis. For example, if we take $m_{S_3} = m_{S_1} = 1.6$\,TeV we find that $|\xi|<3.1 (3.9)$ at the 1\,$\sigma$ (2\,$\sigma$) level. This constraint becomes weaker for heavier LQ masses, as expected from the decoupling limit. For instance, for $m_{S_3} = 1.6$\,TeV and $m_{S_1} = 3$\,TeV the corresponding limit becomes $|\xi|<4.4 (5.6)$ at the 1\,$\sigma$ (2\,$\sigma$) level.

	\item[•]\underline{$\widetilde{S}_1\,\&\, S_3$}: The interactions of $S_3=S_a$ and $\widetilde{S}_1=S_b$ with $H$ are described by
\begin{equation}
\label{eq:mix-S1t-S3}
\mathcal{L}^{\widetilde{S}_1\,\&\, S_3}_{\mathrm{mix}}=\xi\, H^T i \tau_2 (\vec{\tau}\cdot\vec{S_3}) H \widetilde{S}_1^\ast+\mathrm{h.c.}\,,
\end{equation}
which induce a mixing between the $Q=4/3$ component of $S_3$ with $\widetilde{S}_1$, where the states that enter Eq.~\eqref{eq:mix} are $S^{(Q)}_a = S_3^{(4/3)}$ and $S^{(Q)}_b = \widetilde{S}_1^{(4/3)}$. The parameter $\xi$ is normalized in such a way that one gets $\Omega=-\xi \, v^2/2$ in Eqs.~\eqref{eq:mix_matrix} and~\eqref{eq:theta}. The physical masses, in this particular scenario, are $m_{S^{(4/3)}_\pm}$ and $m_{S_3}=m_{S^{(1/3)}_3}=m_{S^{(-2/3)}_3}$. 

The modification of the $T$-parameter for $m_{\widetilde{S}_1} \geq m_{S_3}$ reads 
\begin{equation}
\Delta T_{\widetilde{S}_1\,\&\, {S}_3} = \dfrac{N_c}{8\pi s_W^2}\dfrac{1}{m_W^2}\Bigg{[}\cos^2\theta\, F(m_{S_3},m_{S^{(4/3)}_-})+ \sin^2\theta\, F(m_{S_3},m_{S^{(4/3)}_+})\Bigg{]}\,,
\label{eq:T_2}
\end{equation}

\noindent where the factor of 2 difference between Eqs.~\eqref{eq:T_1} and~\eqref{eq:T_2} comes from the fact that it is the $I^{S_3}_3=+1$ component of $S_3$ that mixes with $\widetilde{S}_1$, while in the $S_1\,\&\, {S}_3$ case the mixing is between the $I^{S_3}_3=0$ component of $S_3$ and $S_1$. The LEP constraint on $\xi$, in the $\widetilde{S}_1\,\&\, S_3$ case, will thus be even weaker than in the $S_1\,\&\, S_3$ case for the LQ masses of interest. For instance, if we take $m_{S_3} = m_{\widetilde{S}_1} = 1.6$\,TeV, the constraint on the mixing parameter reads $|\xi|<4.3 (5.5)$ at the 1\,$\sigma$ (2\,$\sigma$) level, which is weaker than the naive perturbative bound.

	\item[•]\underline{$\widetilde{R}_2\,\&\, R_2$}: The $\widetilde{R}_2\,\&\, R_2$ pair mixes through the operator (see also Ref.~\cite{Kosnik:2012dj}) 
\begin{align}
\label{eq:mix-R2-R2t}
\mathcal{L}^{\widetilde{R}_2\,\&\, R_2}_{\mathrm{mix}} &= - \xi\, \big{(}R_2^\dagger H\big{)}\big{(}\widetilde{R}_2^T i \tau_2 H\big{)}+\mathrm{h.c.}\,.
\end{align}

This interaction yields, after electroweak symmetry breaking, the matrix of Eq.~\eqref{eq:mix_matrix} with $\Omega=-\xi \, v^2/2$, $S^{(Q)}_a = R_2^{(2/3)}$, and $S^{(Q)}_b = \widetilde{R}_2^{(2/3)}$, where we define $S_a = R_2$, $S_b = \widetilde{R}_2$, and note that the states that mix have $Q=2/3$. The masses of four physical states, since $I^{\widetilde{R}_2}=I^{R_2}=1/2$, are $m_{S^{(2/3)}_\pm}$, $m_{\widetilde{R}_2}=m_{\widetilde{R}^{(-1/3)}_2}$, and $m_{R_2}=m_{R^{(5/3)}_2}$, which we assume to satisfy $m_{\widetilde{R}_2}\geq m_{R_2}$.

We find that the $T$-parameter modification induced by the operator of Eq.~\eqref{eq:mix-R2-R2t} is
\begin{align}
\begin{split}
\Delta T_{\widetilde{R}_2\,\&\, R_2} = \dfrac{N_c}{16\pi s_W^2}\dfrac{1}{m_W^2}&\Big{\lbrace}\cos^2\theta\, \Big{[}F(m_{R_2},m_{S^{(2/3)}_-})+F(m_{\widetilde{R}_2},m_{S^{(2/3)}_+})\Big{]} \\
&\qquad\quad+ \sin^2\theta\,\Big{[}F(m_{R_2},m_{S^{(2/3)}_+})+F(m_{\widetilde{R}_2},m_{S^{(2/3)}_-})\Big{]}\Big{\rbrace}\,.
\end{split}
\label{eq:T_3}
\end{align}
For illustration, if we consider $m_{R_2} = m_{\widetilde{R}_2} = 1.6$\,TeV, the $T$-parameter constraint yields $|\xi|<4.3 (5.5)$ at the 1\,$\sigma$ (2\,$\sigma$) level, which is weaker than the naive perturbative limit.

\end{itemize}

With these ingredients at hands, we will now discuss the contributions to $(g-2)_\mu$ in each of the potentially viable scenarios.

\section{$(g-2)_\mu$ via $S_1\,\&\, {S}_3$}
\label{sec:gminus2}

\subsection{Setup}

The $S_1\,\&\, {S}_3$ scenario is the only scenario for which the top quark runs in the $(g-2)_\mu$ loops and it is thus the most promising one of the three. The relevant interactions of $S_1$ and $S_3$ with the SM fermions are given by
\begin{align}
\label{eq:zzza}
\mathcal{L}_{S_1} &= y_R^{ij} \, \overline{u}_{Ri}^C e_{Rj}\, S_1 +\mathrm{h.c.}\,,
\\[0.4em]
\label{eq:zzzb}
\mathcal{L}_{S_3} &= y_L^{ij}\, \overline{Q}^C_i i \tau_2 (\vec{\tau}\cdot \vec{S}_3)L_j+\mathrm{h.c.}\,,
\end{align}
where $y_R$ and $y_L$ are Yukawa coupling matrices and $i$ and $j$ are flavor indices for quarks and leptons, respectively. We omit the $B$ number violating couplings of both $S_1$ and $S_3$ as well as the couplings of $S_1$ with the left-chiral leptons. The latter are omitted since we want to investigate the chirality-enhanced contributions coming solely from the LQ mixing scenario. The Lagrangians in Eqs.~\eqref{eq:zzza} and \eqref{eq:zzzb} can be expanded in terms of the electric charge eigenstates as
\begin{align}
\label{eq:S1S3}
\begin{split}
\mathcal{L}_{S_1 \,\&\, S_3} &= y_R^{ij} \, \bar{u}_{Ri}^C e_{Rj}\, S^{(1/3)}_1 - y_L^{ij}  \bar{d}^{C}_{Li} \nu_{Lj}  \, S_3^{(1/3)}-\sqrt{2} y_L^{ij}  \bar{d}^{\,C}_{Li} e_{Lj}  \, S_3^{(4/3)}\\
&+\sqrt{2} \left(V^*y_L  \right)^{ij}  \bar{u}^C_{Li} \nu_{Lj}  \, S_3^{(-2/3)}-\left(V^* y_L  \right)^{ij}  \bar{u}^C_{Li} e_{Lj}  \, S_3^{(1/3)}  + \text{h.c.}\,,
\end{split}
\end{align}
where $V$ is the Cabibbo-Kobayashi-Maskawa mixing matrix. We take the Pontecorvo-Maki-Nakagawa-Sakata mixing matrix to reside in neutrino sector. In this convention $\nu_{Lj}$ are flavor eigenstates. Note that, after electroweak symmetry breaking, one should replace $S_3^{(1/3)}$ and $S^{(1/3)}_1$ in Eq.~\eqref{eq:S1S3} with $S_{\pm}\equiv S_{\pm}^{(1/3)}$, as defined in Eqs.~\eqref{eq:mix} and~\eqref{eq:mix-S1-S3}, to go to the mass eigenstate basis for electrically charged fields.

\begin{figure}[t!]
\centering
\includegraphics[width=0.75\textwidth]{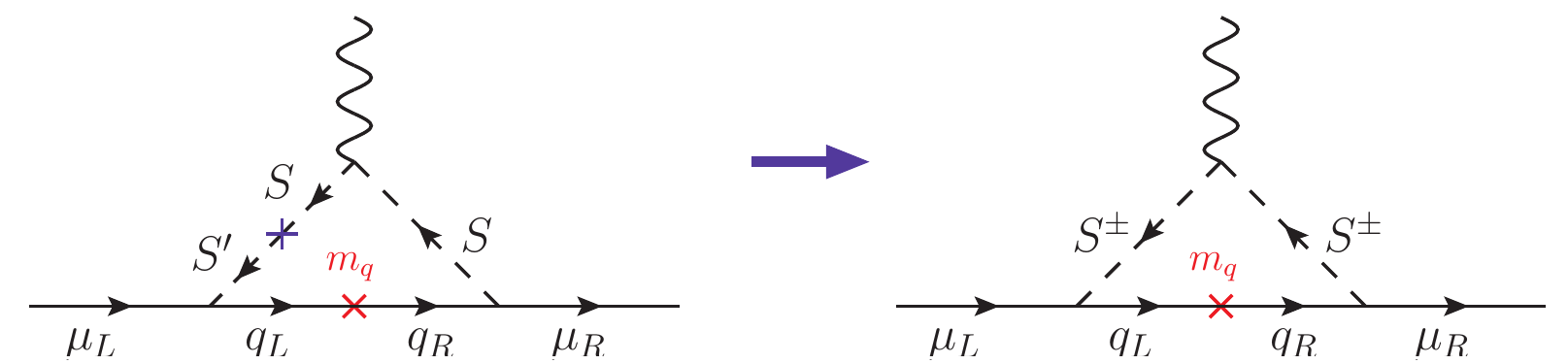}
\caption{Illustration of chirality-enhanced contributions to $(g-2)_\mu$ coming from the mixing of two LQ states $S$ and $S^\prime$  that couple to muons of opposite chiralities. }
\label{fig:diagram-lq-double}
\end{figure}

LQ mixing can induce chirality-enhanced contributions to $(g-2)_\mu$, as illustrated in Fig.~\ref{fig:diagram-lq-double}. To compute $\delta a_\mu$ within this scenario, we replace each of the mass eigenstates running in the loops, i.e., $S_3^{(4/3)}$ and $S_{\pm}$, and their Yukawa interactions in Eq.~\eqref{eq:amu}. Note that we assume that $y_R^{t\mu}$ and $y_L^{b\mu}$, as defined in Eqs.~\eqref{eq:zzza} and \eqref{eq:zzzb}, are the only non-zero entries in $y_R$ and $y_L$. The final result reads
\begin{align}
\label{eq:amu-S1-S3full}
\delta a_{\mu}= -\dfrac{N_c\,m_\mu^2}{8\pi^2}\Bigg{\lbrace}2\, |y_L^{b\mu}|^2\,\dfrac{ \mathcal{F}_{4/3}(x_b)}{m_{S_3}^2}&+\left[\sin^2 \theta\, |y_R^{t\mu}|^2+\cos^2 \theta \,|y_L^{b\mu}|^2\right]\dfrac{\mathcal{F}_{1/3}(x_t^-)}{m_{S_-}^2}\\
&+\left[{\cos^2 \theta \,|y_R^{t\mu}|^2+\sin^2 \theta\, |y_L^{b\mu}|^2}\right]\dfrac{\mathcal{F}_{1/3}(x_t^+)}{m_{S_+}^2}\nonumber\\
&+\dfrac{m_t} {m_\mu}\sin \theta\cos\theta \,\mathrm{Re}\big{(}y_L^{b\mu}\,y_R^{t\mu\,\ast}\big{)}\Bigg{[}\dfrac{\mathcal{G}_{1/3}(x_t^+)}{m_{S_+}^2}-\dfrac{\mathcal{G}_{1/3}(x_t^-)}{m_{S_-}^2}\Bigg{]}\Bigg{\rbrace}\,,\nonumber
\end{align}
where $x_b=m_b^2/{m_{S_3}^2}$ and $x_t^{\pm}=m_t^2/{m_{S_\pm}^2}$. These expressions have been further simplified by taking $|V_{tb}|\approx 1$. The loop functions $\mathcal{F}_Q$ and $\mathcal{G}_Q$ are defined in Eqs.~\eqref{eq:F-Q} and \eqref{eq:G-Q}. In the absence of LQ mixing, the chirality-enhanced contribution in the last line would vanish since, in that instance, $m_{S_+}=m_{S_-}$.

\subsection{Phenomenological constraints}
\label{ssec:constraints}

We shall now discuss the constraints on the LQ Yukawa couplings. Again, to obtain the largest possible contribution to $\delta a_\mu$, coming from top quark in the loops, we assume that $y_R^{t\mu}$ and $y_L^{b\mu}$ of Eqs.~\eqref{eq:zzza} and~\eqref{eq:zzzb} are the only non-zero Yukawa couplings. With this ansatz, most of the low-energy observables are unaffected at first approximation by the LQ presence. The only exceptions are the $Z$-pole observables and the LHC signatures, as we describe below.

\begin{itemize}
\item[•] \underline{$Z\to \ell\ell$}:  
LQs contribute to the $Z$-boson couplings to leptons at the one-loop level. These contributions can be generically parameterized as
\begin{equation}
\mathcal{L}^Z_{\mathrm{eff}} = \dfrac{g}{c_W}\sum_{f,i,j} \bar{f}_i \gamma^\mu \Big{[}g_{f_L}^{ij}\, P_L + g_{f_R}^{ij}\,P_R\Big{]}f_j\, Z_\mu\,,
\end{equation}
where $g$ is the $SU(2)_L$ gauge coupling and $
g_{f_L(R)}^{ij} = \delta_{ij}\, g_{f_{L(R)}}^{\mathrm{SM}}+\delta g_{f_{L(R)}}^{ij}$ with $g_{f_L}^{\mathrm{SM}}=I_3^f-Q^f s_W^2$ and $g_{f_R}^{\mathrm{SM}}=-Q^f\, s_W^2$, while $\delta g_{f_{L(R)}}^{ij}$ contains the LQ contributions. In these expressions, $I_3^f$ denotes the third weak-isospin component and $Q_f$ denotes the electric charge of a fermion $f\in \lbrace u,d,\ell,\nu \rbrace$. The contributions arising separately from $S_1$ and $S_3$ states to $\delta g_{f_{L(R)}}^{ij}$ have been computed in Ref.~\cite{Arnan:2019olv} at the one-loop order. For completeness, we spell out the results from Ref.~\cite{Arnan:2019olv} in the following,
\begin{align}
\label{eq:deltagl}
\delta g_{\ell_{R}}^{22} &= \dfrac{N_C\,|y^{t\mu}_R|^2}{32\pi^2}\bigg{[}-\dfrac{x_t(x_t-1-\log x_t)}{(x_t-1)^2}+ \dfrac{x_Z}{6}F_2^{\ell_R}(x_t)\bigg{]} \,,\\[0.5em]
\begin{split}
\label{eq:deltagnu}
\delta g_{\ell_{L}}^{22} &= \dfrac{N_C\,|y^{b\mu}_L|^2}{32\pi^2}\bigg{[}\dfrac{x_t(x_t-1-\log x_t)}{(x_t-1)^2}+ \dfrac{x_Z}{6}F_2^{\ell_L}(x_t)\bigg{]}\\[0.3em]
&\hspace*{5em}+ \dfrac{N_C \, x_Z\,|y_L^{b\mu}|^2}{24\pi^2}\bigg{[}g_{d_R}^{\mathrm{SM}} \bigg{(}\log x_Z - i \pi - \dfrac{1}{6}\bigg{)}+\dfrac{g_{\ell_L}^\mathrm{SM}}{6}\bigg{]}\,,
\end{split} 
\end{align}

\noindent where we have replaced $|V_{tb}|\approx 1$ to simplify the second expression, and denoted $x_t=m_t^2/m_S^2$ and $x_Z= m_Z^2/m_S^2$, by setting $m_S \approx m_{S_1} \approx m_{S_3}$. The latter approximation is valid since the mass splitting between $m_{S^{\pm}}$ and $m_{S_{1(3)}}$ is already tigtly constrained by the $T$-parameter, entailing a negligible contribution to the $Z$-boson couplings. The coupling $\delta g_{\nu_{L}}^{22}$ can be obtained from Eq.~\eqref{eq:deltagnu} by replacing $\ell_L \to \nu_L$. The functions $F_2^{\ell_L(\ell_R)}$ account for $\mathcal{O}(m_Z^2/m_t^2)$ corrections to the effective couplings, which are given by~\cite{Arnan:2019olv} 
\begin{align}
\begin{split}
F_{2}^{\ell_{L}(\ell_R)}(x_t) = &-g_{u_{L(R)}}^\mathrm{SM} \dfrac{(x_t-1)(5 x_t^2-7 x_t+8)-2(x_t^3+2)\log x_t}{(x_t-1)^4} \\&- g_{u_{R(L)}}^\mathrm{SM} \dfrac{(x_t-1)(x_t^2-5 x_t-2)+ 6 x_t\log x_t}{(x_t-1)^4}  \\&+ g_{\ell_{L(R)}}^\mathrm{SM} \dfrac{(x_t-1)(-11 x_t^2+ 7 x_t-2)+ 6 x_t^3\log x_t}{3(x_t-1)^4}  \,.
\end{split}
\end{align}
\noindent For the experimental inputs, we consider the fit to LEP data performed in Ref.~\cite{Efrati:2015eaa}, which includes the correlation between the $W$ and $Z$ boson couplings to leptons and neutrinos coming from gauge invariance, amounting to the following  $2\,\sigma$ constraints on the effective couplings of the $Z$-boson to muons,
\begin{align}
\begin{split}
\delta g_{\ell_R}^{22} &\in (-1.5,0.6)\times 10^{-3}\,, \qquad\quad \delta g_{\ell_L}^{22} \in (-0.5,1.0)\times 10^{-3}\,,\\[0.5em] \qquad \delta g_{\nu_L}^{22} &\in (-0.9,0.15)\times 10^{-2}\,,
\end{split}
\end{align}

\noindent which are considered, along with their correlation, in our analyses.~\footnote{For comparison, these constraints are compatible to the ones arising solely from the $Z$-boson width into leptons~\cite{Feruglio:2016gvd}. See also Ref.~\cite{Coy:2019rfr} for a recent discussion in the context of the SM effective field theory.
}

\item[•] \underline{The LHC searches}: LQ $S$ can be pair-produced in hadron colliders via $gg(q\bar{q})\to S^* S$. These pairs can be searched for via subsequent decay into quark-lepton pairs. See, for example, Refs.~\cite{Diaz:2017lit,Angelescu:2018tyl} for recent reviews on the subject. For the scenarios we consider, the most stringent bounds come from direct searches for scalar LQs decaying into $b\mu$ ($t\mu$) final states, setting the limit $m_S \gtrsim 1400$\,GeV ($m_S \gtrsim 1420$\,GeV) for $\mathcal{B}(S\to b\mu)=1$~\cite{Diaz:2017lit} ($\mathcal{B}(S\to t\mu)=1$~\cite{Angelescu:2018tyl}). If we considered instead an LQ coupling to second generation quarks, the limit obtained would be $m_S \gtrsim 1530$\,GeV for $\mathcal{B}(S\to j \mu)=1$~\cite{CMS:2018sgp}. In our phenomenological analysis we are mainly focused on the scenarios that introduce four, for all practical purposes, mass degenerate LQ states, some of which certainly have the same decay signatures to final states with muons. We use the approach advocated in Ref.~\cite{Diaz:2017lit} to accommodate this feature and deduce that the conservative mass limit, under the assumption that $\mathcal{B}(S\to j\mu)=1$, where $S$ is the relevant LQ and $j$ is the relevant quark, and using experimental input of Refs.~\cite{ATLAS:2017hbw,CMS:2017iir}, is $m_S \gtrsim 1.6$\,TeV. We use this limit even when we consider non-degenerate LQ spectrum, in what follows. 

LQ couplings can also be indirectly probed via the study of high-$p_T$ dilepton-tails at LHC~\cite{Greljo:2017vvb}. In this way, one can constrain, for instance, the LQ coupling $y_L^{b\mu}$ as a function of LQ mass. In our analysis, we consider the constraints obtained in Ref.~\cite{Angelescu:2018tyl} for different LQ states with couplings to muons from $36\,\mathrm{fb}^{-1}$ of LHC data and offer projection for the $300\,\mathrm{fb}^{-1}$ data set.
\end{itemize}

\subsection{Numerical results}

\begin{figure}[!t]
\centering
\includegraphics[width=0.49\textwidth]{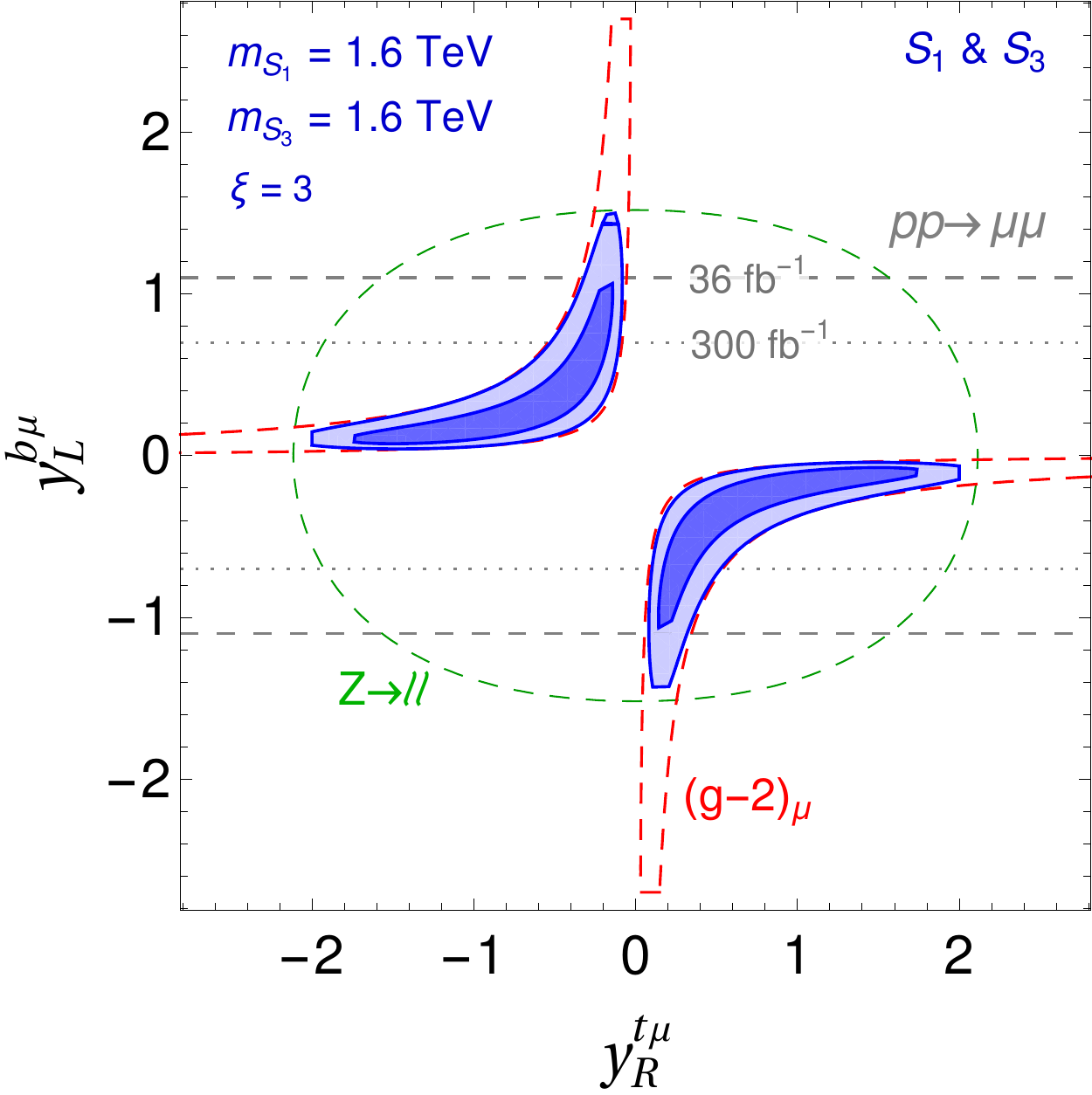}
\includegraphics[width=0.49\textwidth]{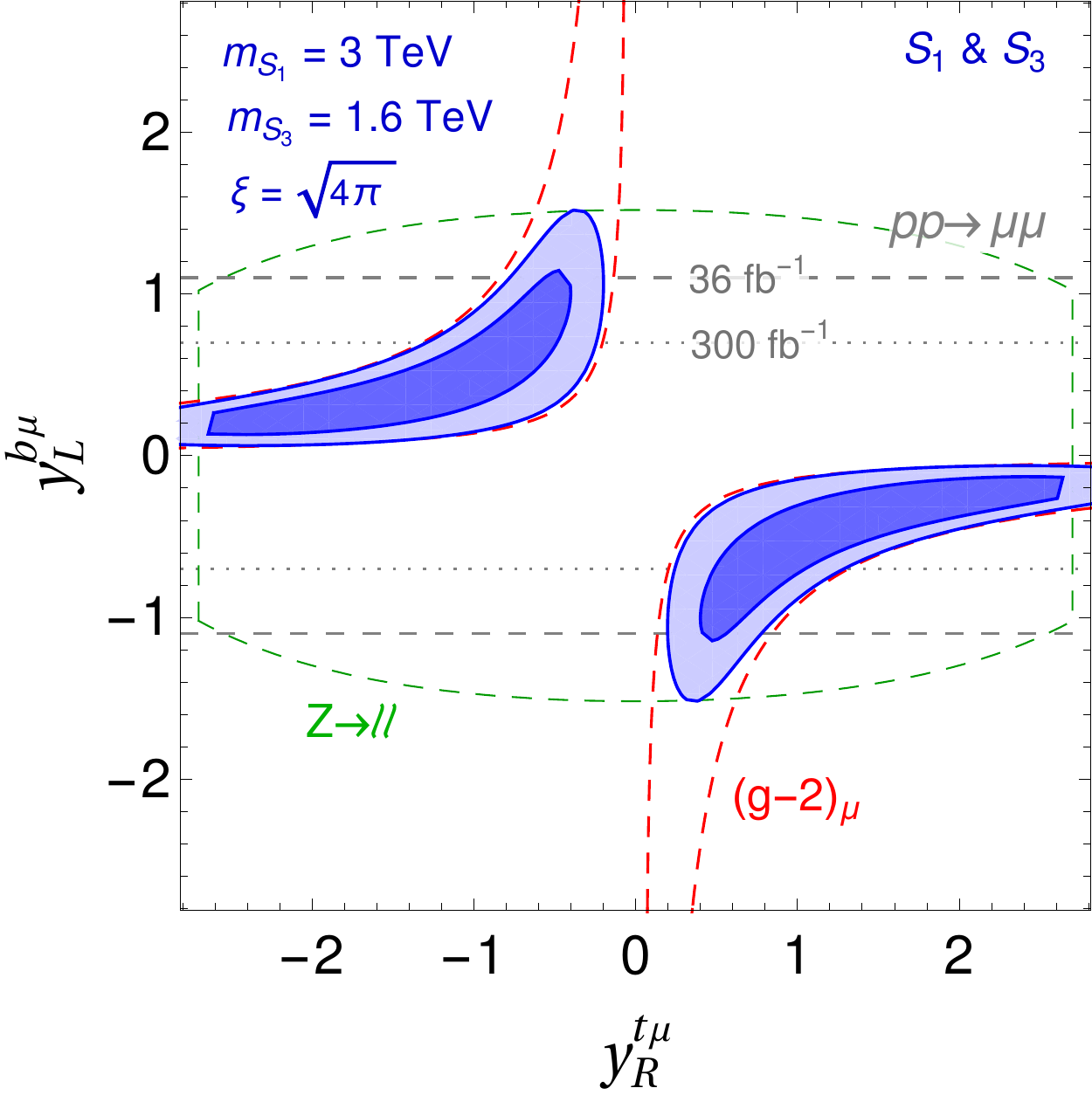}
\caption{The values of $y_R^{t\mu}$ vs.\ $y_L^{b\mu}$ for the $S_1\,\&\, {S}_3$ scenario that satisfy relevant flavor constraints and address $(g-2)_\mu$ through the top-quark loops are shown in dark (light) blue to $1\,\sigma$ ($2\,\sigma$) accuracy for the mass degenerate case with $\xi =3$ and mass non-degenerate case with $\xi =\sqrt{4 \pi}$ in the left and right panels, respectively. The $2\,\sigma$ constraints from the $Z$-pole observables and $(g-2)_\mu$ are represented with green and red dashed lines, respectively. The (projected) LHC limits from the study of $pp\to \mu\mu$ at high-$p_T$, at the $2\,\sigma$ level, for ($300$\,fb$^{-1}$) $36$\,fb$^{-1}$ worth of data are shown with (dash-dotted) dashed gray lines~\cite{Angelescu:2018tyl}.}
\label{fig:plot-S1-S3}
\end{figure}

There are five parameters that are relevant for our numerical analysis. These are
\begin{equation}
\lbrace m_{S_3}, m_{S_1}, \xi, y_L^{b\mu}, y_R^{t\mu} \rbrace
\end{equation}
that are defined through Eqs.~\eqref{eq:mix-S1-S3},~\eqref{eq:zzza}, and~\eqref{eq:zzzb}. We first consider the scenario with the maximal mixing by setting $ m_{S_3} = m_{S_1} = 1.6$\,TeV and with the top quark in the $(g-2)_\mu$ loops. We furthermore choose $\xi =3$ that is still in agreement with the bounds obtained through Eq.~\eqref{eq:T_1} and we impose the constraints from Sec.~\ref{ssec:constraints} onto the LQ Yukawa couplings. The parameter space of $y_L^{b\mu}$ vs.\ $y_R^{t\mu}$ that can address $(g-2)_\mu$ is then presented in the left panel of Fig.~\ref{fig:plot-S1-S3}. The scenario when $\xi =\sqrt{4 \pi}$, $m_{S_3} = 1.6$\,TeV and $m_{S_1} = 3$\,TeV is shown in the right panel of Fig.~\ref{fig:plot-S1-S3}. Note that, in the latter scenario, the $T$-parameter constraint turns out to be weaker than the naive perturbative bound, i.e., $|\xi|\leq \sqrt{4\pi}$, due to the large LQ mass, as shown in Sec.~\eqref{sec:ewpt}. If one is interested in the allowed parameter space of $y_L^{b\mu}$ vs.\ $y_R^{t\mu}$ for any other $\xi$ all one needs to do is to keep the product $(\xi \,y_L^{b\mu} y_R^{t\mu})$, as given in any part of the viable parameter space of Fig.~\ref{fig:plot-S1-S3}, at the associated value. 

\begin{figure}[t!]
\centering
\includegraphics[width=0.55\textwidth]{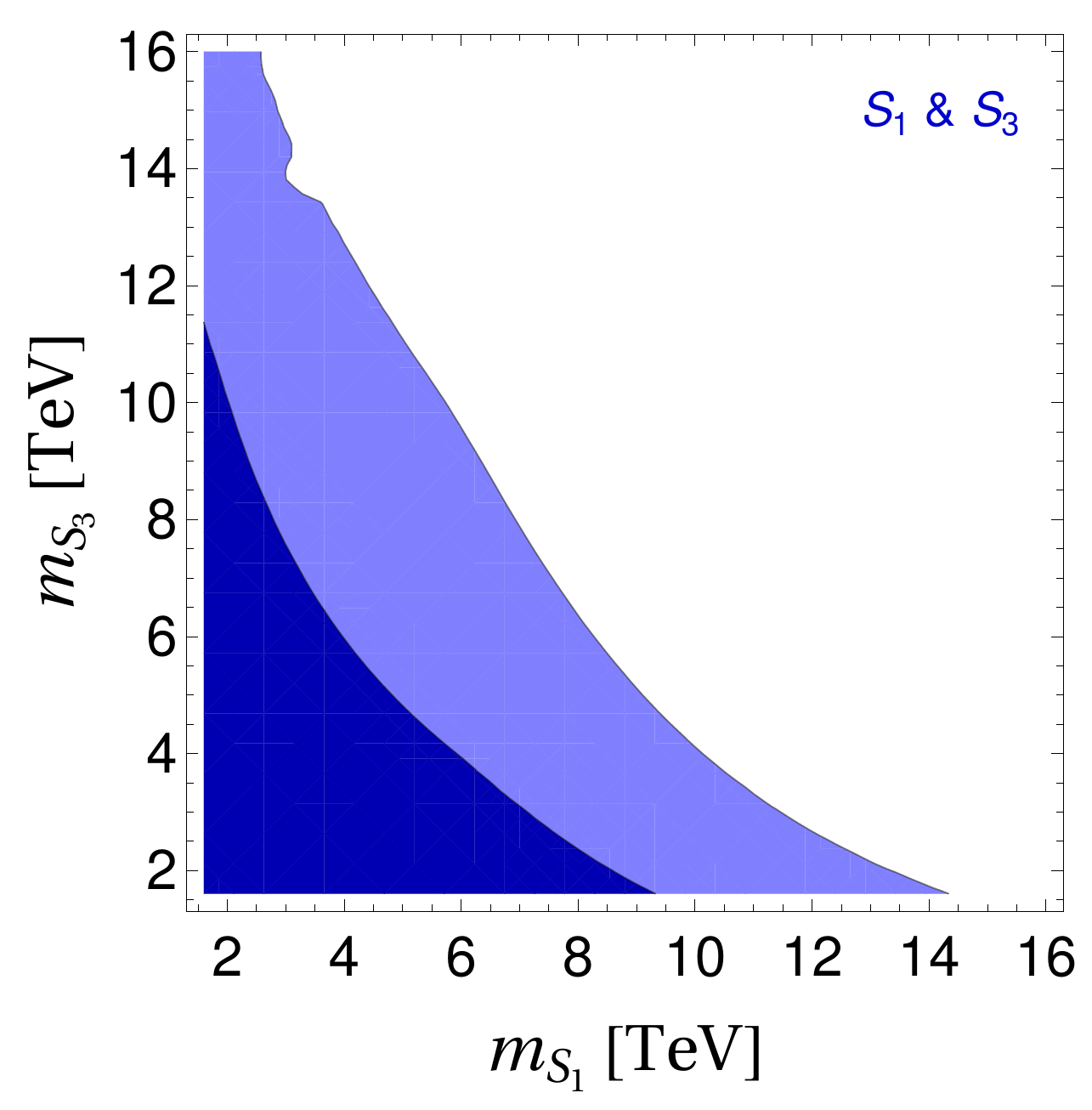}
\caption{Allowed values of $m_{S_1}$ and $m_{S_3}$ that can address $(g-2)_\mu$, while being consistent with other flavor-physics constraints, at the 1\,$\sigma$ (2\,$\sigma$) level in dark (light) blue. We conservatively take $|y_L^{b\mu}|,|y_R^{t\mu}|, |\xi| \lesssim \sqrt{4\pi}$. }
\label{fig:plot-S1-S3-general}
\end{figure}

Next, we allow both $m_{S_1}$ and $m_{S_3}$ to vary in order to find their maximal values that are still consistent with the perturbative bound on Yukawa couplings and $\xi$, where we conservatively take $|y_L^{b\mu}|,|y_R^{t\mu}|, |\xi| \leq \sqrt{4\pi}$. The correlation between $m_{S_1}$ and $m_{S_3}$ is shown in Fig.~\ref{fig:plot-S1-S3-general}. Clearly, the maximal allowed common mass scale that can accommodate $(g-2)_\mu$ at the 1\,$\sigma$ level, while remaining consistent with other flavor bounds, is approximately at $5$\,TeV, which is possibly within the reach of high-energy LHC. Note that this limit is considerably lower than the one obtained for the single LQ solutions of the $(g-2)_\mu$ discussed in Sec.~\ref{sec:single-lq}, for which the maximal LQ mass is of the order of $100$\,TeV~\cite{Queiroz:2014zfa,Biggio:2014ela,Bauer:2015knc,ColuccioLeskow:2016dox,Kowalska:2018ulj,Mandal:2019gff}.

\subsection{Can charm work in $S_1 \,\&\, S_3$ scenario?}

We briefly discuss the viability of a scenario where the leading contribution to $(g-2)_\mu$ comes not from the top-quark loops, but from the charm-quark loops instead. In that case, the chiral enhancement is not as significant as it is in the top quark case but could be, in principle, still large enough to significantly impact $\delta a_\mu$. As we are going to demonstrate, this naive reasoning is not correct due to additional constraints from flavor physics that become relevant in this scenario. 

We assume that the only non-zero LQ couplings in Eqs.~\eqref{eq:zzza} and~\eqref{eq:zzzb} are $y_R^{c\mu}$ and $y_L^{s\mu}$, and we focus on the maximal-mixing scenario, with $m_S\equiv m_{S_1}=m_{S_3}$. By keeping only the dominant, chirality enhanced, contributions to $\delta a_\mu$, we find that 
\begin{equation}
\delta a_\mu \approx -3 \times 10^{-9}\,\left(\dfrac{\xi}{1.9}\right)\,\left( \dfrac{\mathrm{Re}\big{(} y_L^{s\mu}\, y_R^{c\mu}\big{)}}{1.5}\right)\,\left(\dfrac{1\,\mathrm{TeV}}{m_S}\right)^4\,,
\label{eq:amu-charm}
\end{equation}
where $|\xi|<1.9$ corresponds to the $T$-parameter constraint for $m_S=1$\,TeV to $1\,\sigma$ accuracy. From this expression, we see that the description of $a_\mu^{\mathrm{exp}}$ can be improved for $m_S \approx 1$\,TeV and $\mathcal{O}(1)$ Yukawa couplings, in agreement with the LHC data and perturbativity limits. The main constraint, which actually precludes this possibility, comes from charm physics. Namely, the LQ mixing also induces a contribution to rare charm decays, lifting the SM helicity suppression as follows,
\begin{align}
\begin{split}
\mathcal{B}(D^0\to\mu\mu) &\approx \dfrac{\tau_{D_0}f_D^2 m_{D^0}^3}{1024 \pi} \left(\dfrac{\xi\,v^2}{m_S^4}\right)^2 \left|y_R^{c\mu}\,(V_{us}\, y_L^{s\mu})^\ast\right|^2 \left(\dfrac{m_{D^0}}{m_c}\right)^2 \beta_\mu\left(1-\dfrac{2 m_\mu^2}{m_{D^0}^2} \right) \,,
\end{split}
\end{align}
where we keep only the dominant LQ contribution, define $\beta_\mu=\sqrt{1- {4m_\mu^2}/{m_{D^0}^2}}$, and take $f_D=209(3)$\,MeV~\cite{Aoki:2019cca} to be $D$-meson decay constant. Since the current experimental limit reads $\mathcal{B}(D^0\to\mu\mu)^{\mathrm{exp}}<6.2\times 10^{-9}$~\cite{Aaij:2013cza}, we find, for the maximal allowed value of $\xi$ when $m_S\approx 1$\,TeV, that
\begin{equation}
\dfrac{|y_L^{s\mu} y_R^{c\mu}|}{m_S^2} \lesssim 0.12\,\mathrm{TeV}^{-2}\,.
\end{equation}
Therefore, the LQ contribution, as defined in Eq.~\eqref{eq:amu-charm}, can only marginally contribute towards the current value of $\delta a_\mu$.

\section{$(g-2)_\mu$ via $\widetilde{S}_1\,\&\, S_3$}
\label{sec:others_2}

The $\widetilde{S}_1\,\&\, S_3$ scenario is rather special since neither of these LQ states can explain the $(g-2)_\mu$ discrepancy separately. In other words, the only way they can generate chirality-enhanced contributions towards the anomalous magnetic moment of muon is if they mix through the interactions with the SM Higgs field. 

\begin{figure}[t!]
\centering
\includegraphics[width=0.49\textwidth]{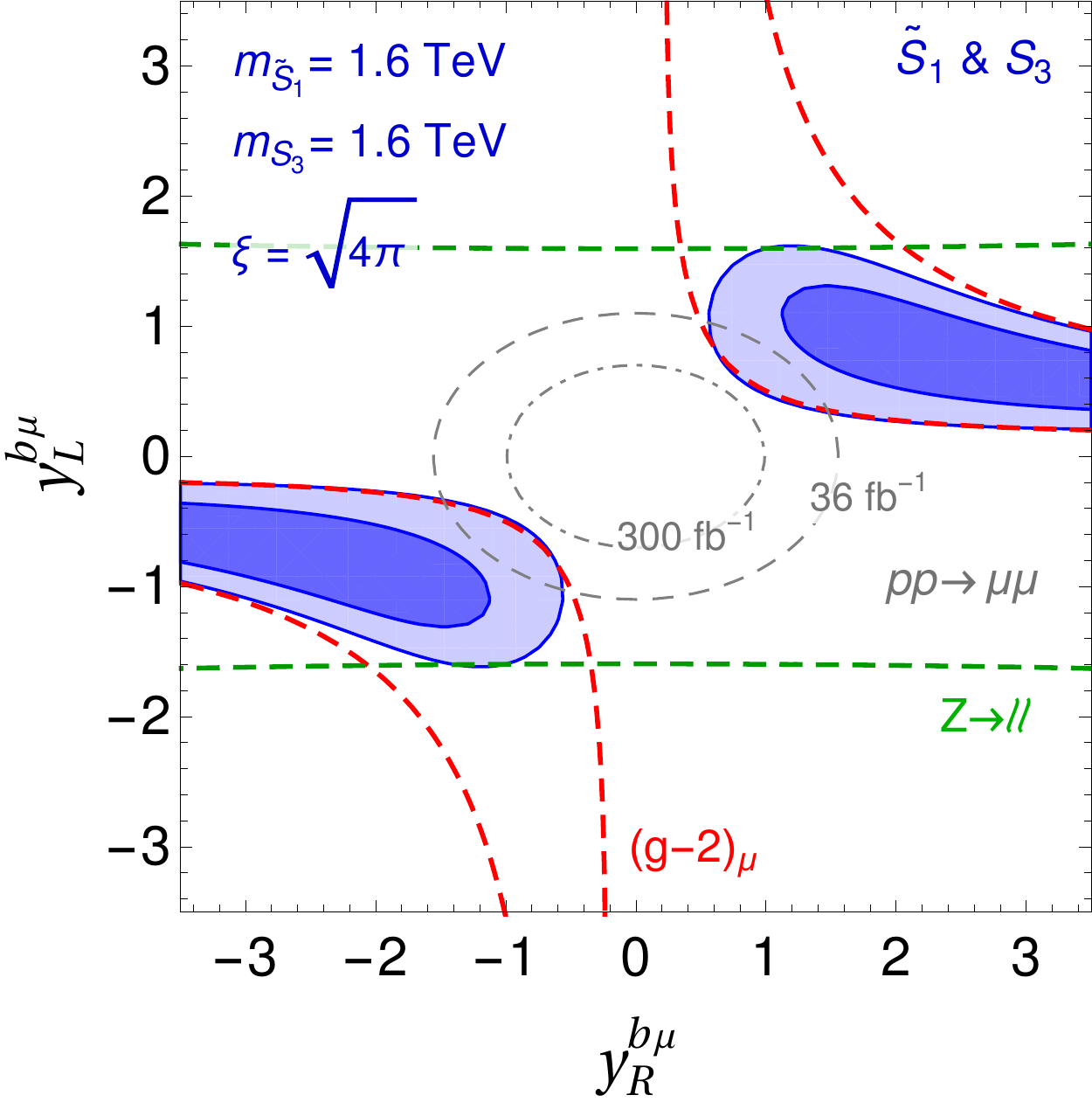}
\includegraphics[width=0.49\textwidth]{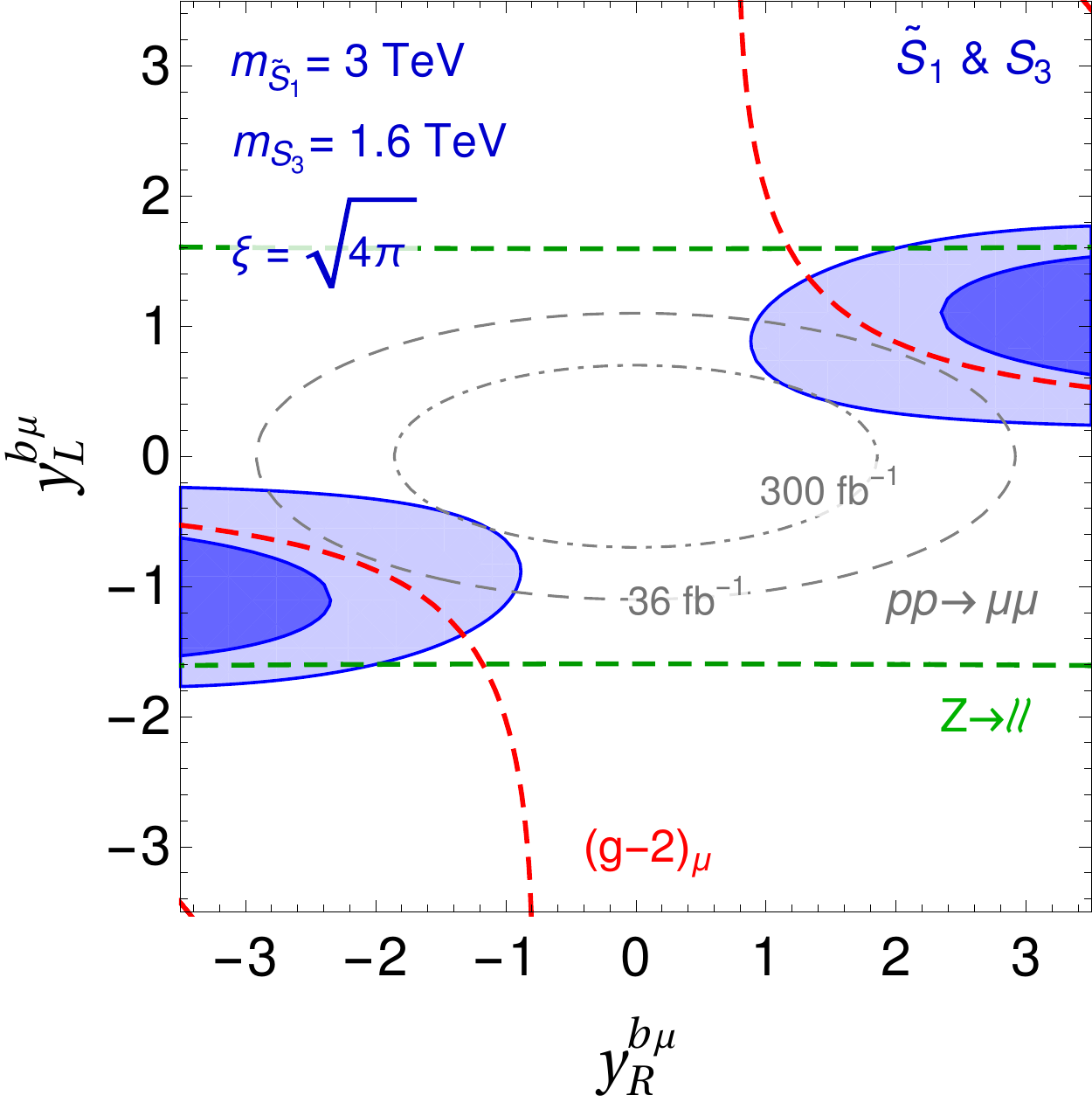}
\caption{The values of $y_R^{b\mu}$ vs.\ $y_L^{b\mu}$ for the $\widetilde{S}_1\,\&\, S_3$ scenario with $\xi =\sqrt{4 \pi}$ that satisfy relevant flavor constraints and address $(g-2)_\mu$ through the bottom-quark loops are shown in dark (light) blue to $1\,\sigma$ ($2\,\sigma$) accuracy for the mass degenerate and mass non-degenerate cases in the left and right panels, respectively. The individual $2\,\sigma$ constraints from the $Z$-pole observables and $(g-2)_\mu$ are represented with green and red dashed lines, respectively. The (projected) LHC limits from the study of $pp\to \mu\mu$ at high-$p_T$, at the $2\,\sigma$ level, for ($300$\,fb$^{-1}$) $36$\,fb$^{-1}$ worth of data are shown with (dash-dotted) dashed gray lines~\cite{Angelescu:2018tyl}.}
\label{fig:S1tilde-final}
\end{figure}

The most dominant contribution towards $(g-2)_\mu$ in the $\widetilde{S}_1\,\&\, S_3$ scenario, according to Table~\ref{tab:01}, comes from the bottom quark in the loops of Fig.~\ref{fig:diagram-lq-double}. To demonstrate that, we spell out relevant interactions of $\widetilde{S}_1$ and $S_3$ with the SM fermions. These are 
\begin{align}
\label{eq:aaa}
\mathcal{L}_{\widetilde{S}_1} &= y_R^{ij}\, \bar{d}_{Ri}^C\,e_{Rj} \,\widetilde{S}_1+\mathrm{h.c.}\,,
\\[0.4em]
\label{eq:bbb}
\mathcal{L}_{S_3} &= y_L^{ij}\, \bar{Q}^C_i i \tau_2 (\vec{\tau}\cdot \vec{S}_3)L_j+\mathrm{h.c.}\,,
\end{align}
where $y_R$ and $y_L$ are Yukawa coupling matrices. We omit the $B$ number violating couplings of both $\widetilde{S}_1$ and $S_3$. The Lagrangians in Eqs.~\eqref{eq:aaa} and~\eqref{eq:bbb} can be expanded in terms of the electric charge eigenstates as
\begin{align}
\begin{split}
\mathcal{L}_{\widetilde{S}_1\,\&\, S_3} &= y_R^{ij}\, \bar{d}_{Ri}^C\,e_{Rj} \,\widetilde{S}^{(4/3)}_1 - y_L^{ij}  \bar{d}^{C}_{Li} \nu_{Lj}  \, S_3^{(1/3)}-\sqrt{2} y_L^{ij}  \bar{d}^{\,C}_{Li} e_{Lj}  \, S_3^{(4/3)}\\
&+\sqrt{2} \left(V^*y_L  \right)^{ij}  \bar{u}^C_{Li} \nu_{Lj}  \, S_3^{(-2/3)}-\left(V^* y_L  \right)^{ij}  \bar{u}^C_{Li} e_{Lj}  \, S_3^{(1/3)}  + \text{h.c.}\,,
\end{split}
\label{eq:S1tS3}
\end{align}
where we see that, in order to close the loops of Fig.~\ref{fig:diagram-lq-double}, we need to mix the $Q=4/3$ components of $\widetilde{S}_1$ and $S_3$, as accomplished by the operator introduced in Eq.~\eqref{eq:mix-S1t-S3}. There are four LQ mass eigenstates in this particular scenario, as discussed in Sec.~\ref{sec:lq-mixing}, with the masses $m_{S_\pm} \equiv m_{S^{(4/3)}_\pm}$ and $m_{S_3} \equiv m_{S_3^{(1/3)}} = m_{S_3^{(-2/3)}}$, where we define the mass eigenstates for mixed states $\widetilde{S}^{(4/3)}_1$ and $S^{(4/3)}_3$ as $S_{\pm} \equiv S_{\pm}^{(4/3)}$. 

We proceed following the same steps that are used to analyse the $S_1\,\&\, {S}_3$ scenario and accordingly switch on only those Yukawa couplings in Eq.~\eqref{eq:S1tS3} that are absolutely necessary to generate contribution towards $(g-2)_\mu$ via the bottom quark exchange. These are $y_R^{b\mu}$ and $y_L^{b\mu}$, where we set all other entries in $y_R$ and $y_L$ of Eqs.~\eqref{eq:aaa} and~\eqref{eq:bbb} to zero. This time the five relevant parameters for our numerical analysis are $m_{S_3}$, $m_{\widetilde{S}_1}$, $\xi$, $y_L^{b\mu}$, and $y_R^{b\mu}$.

The resulting analysis for $m_{S_3} = m_{\widetilde{S}_1} \equiv 1.6$\,TeV yields phenomenologically viable parameter space in the $y_L^{b\mu}$-$y_R^{b\mu}$ plane that we show in the left panel of Fig.~\ref{fig:S1tilde-final}. The corresponding analysis for the case when $m_{S_3} = 1.6$\,TeV and $m_{\widetilde{S}_1} = 3$\,TeV is shown in the right panel of Fig.~\ref{fig:S1tilde-final}. In both instances we take $\xi=\sqrt{4 \pi}$, which is consistent with electroweak-precision constraints. Again, one can change the value of $\xi$ as long as the product $(\xi \,y_L^{b\mu}y_R^{b\mu})$, as given in Fig.~\ref{fig:S1tilde-final}, is kept at the associated value to determine corresponding parameter space.

It is clear from Fig.~\ref{fig:S1tilde-final} that in order to accommodate $(g-2)_\mu$ within the $\widetilde{S}_1\,\&\, S_3$ scenario one needs rather large Yukawa couplings. Also, the LHC limits from the study of $pp\to \mu\mu$ at high-$p_T$ only leave a sliver of otherwise viable parameter space and any improvement of aforementioned limits would significantly impair the ability of the $\widetilde{S}_1\,\&\, S_3$ scenario to address the $(g-2)_\mu$ discrepancy. 

\section{$(g-2)_\mu$ via $\widetilde{R}_2\,\&\, R_2$}
\label{sec:others_1}

\begin{figure}[t!]
\centering
\includegraphics[width=0.49\textwidth]{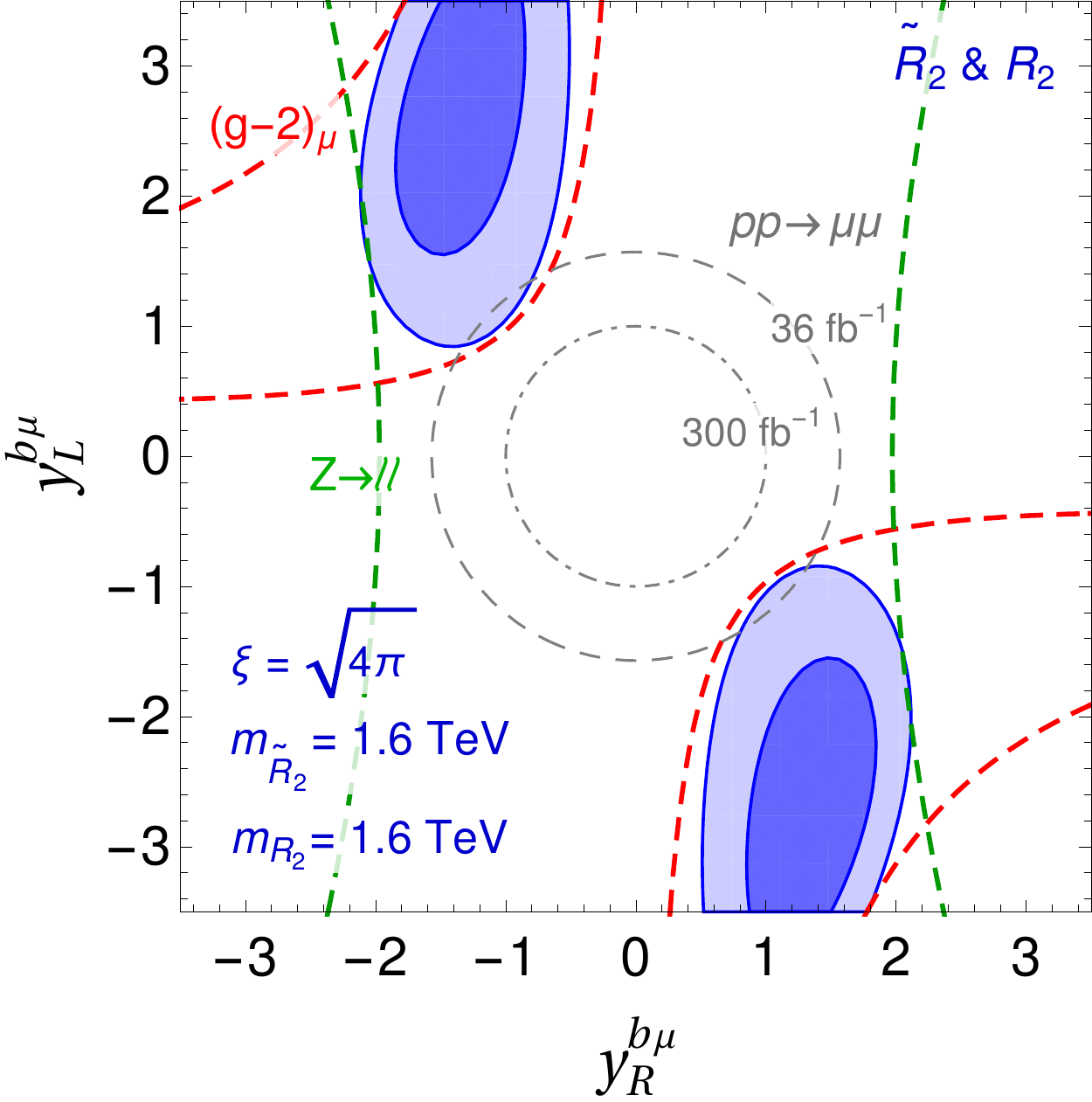}
\includegraphics[width=0.49\textwidth]{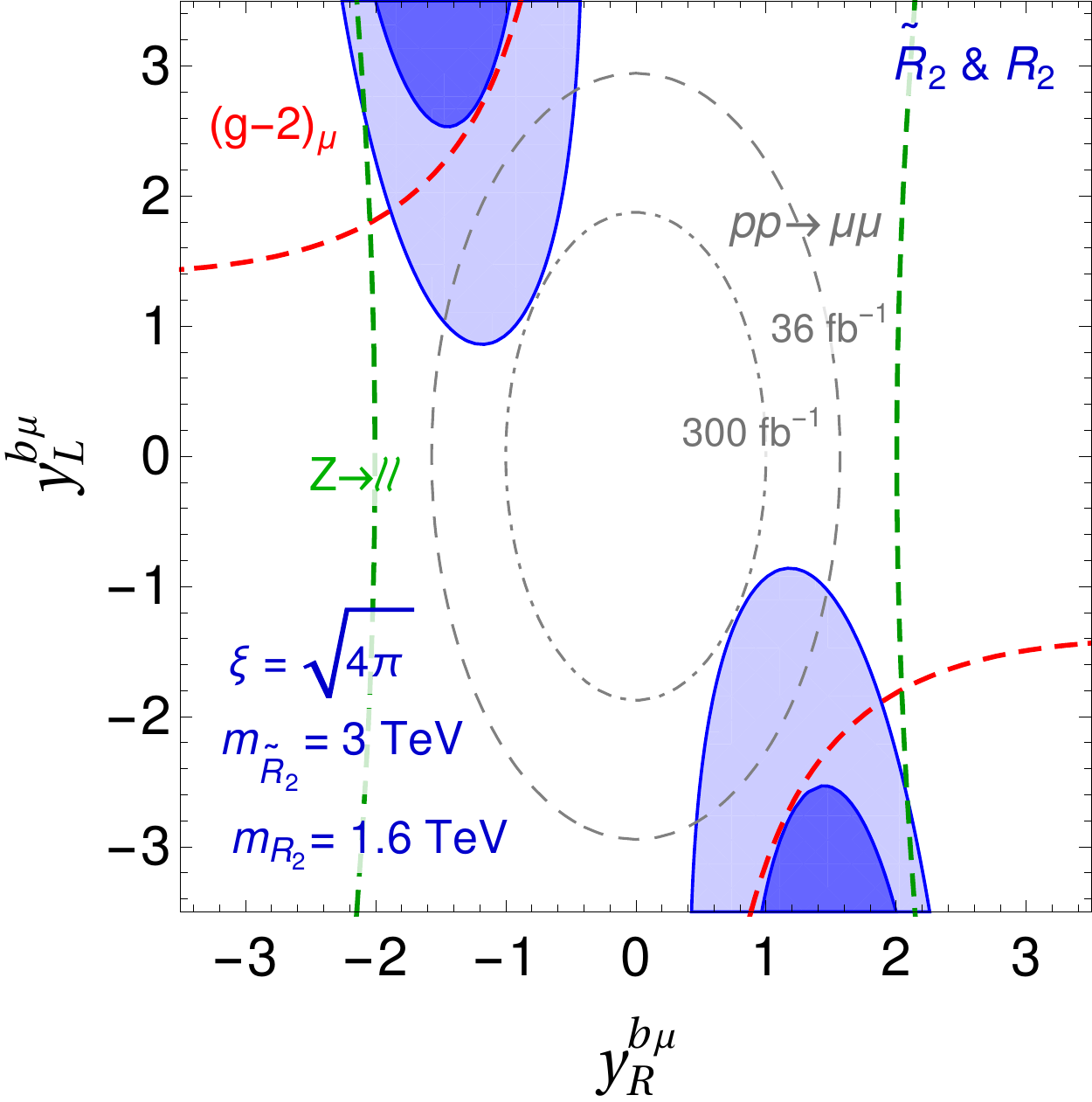}
\caption{The values of $y_R^{b\mu}$ vs.\ $y_L^{b\mu}$ for the $\widetilde{R}_2\,\&\, R_2$ scenario that satisfy relevant flavor constraints and address $(g-2)_\mu$ are shown in dark (light) blue to $1\,\sigma$ ($2\,\sigma$) accuracy. The individual $2\,\sigma$ constraints from the $Z$-pole observables and $(g-2)_\mu$ are represented with green and red dashed lines, respectively. The (projected) LHC limits from the study of $pp\to \mu\mu$ at high-$p_T$, at the $2\,\sigma$ level, for ($300$\,fb$^{-1}$) $36$\,fb$^{-1}$ worth of data are shown with (dash-dotted) dashed gray lines~\cite{Angelescu:2018tyl}.}
\label{fig:R2-final}
\end{figure}

The dominant contribution towards $(g-2)_\mu$ in the $\widetilde{R}_2\,\&\, R_2$ scenario comes from the bottom quark in the loops of Fig.~\ref{fig:diagram-lq-double}. The relevant interactions of $\widetilde{R}_2$ and $R_2$ with the SM fermions are
\begin{align}
\label{eq:sss}
\mathcal{L}_{\widetilde{R}_2} &= -y_L^{ij} \, \overline{d}_{Ri} \widetilde{R}_2 i \tau_2 L_j+\mathrm{h.c.}\,,\\[0.35em]
\label{eq:ttt}
\mathcal{L}_{R_2}&= y_{R}^{ij}\, \overline{Q}_i e_{Rj} R_2+\mathrm{h.c.}\,,
\end{align}
where $y_L$ and $y_R$ are Yukawa coupling matrices. We omit the couplings of $R_2$ with right-chiral leptons to investigate the chirality-enhanced contributions coming solely from the LQ mixing scenario. The above Lagrangians can be expanded in terms of charge eigenstates as
\begin{align}
\label{eq:R2tR2}
\begin{split}
\mathcal{L}_{\widetilde{R}_2\,\&\, R_2} = &-y_L^{ij}\bar{d}_{Ri}e_{Lj}\widetilde{R}_{2}^{(2/3)}+y_L^{ij}\bar{d}_{Ri}\nu_{Lj}\widetilde{R}_{2}^{(-1/3)}\\
&+(V y_R)^{ij} \bar{u}_{Li} 
  e_{Rj} R_2^{(5/3)} +y_R^{ij} \bar{d}_{Li} e_{Rj} R_2^{(2/3)}+\mathrm{h.c.}\,,
  \end{split}
\end{align}
where, in order to close the loops of Fig.~\ref{fig:diagram-lq-double}, we need to mix the $Q=2/3$ components of $\widetilde{R}_2$ and $R_2$, as achieved via the operator introduced in Eq.~\eqref{eq:mix-R2-R2t}. After the Higgs field gets the vacuum expectation value we are left with four mass eigenstates with masses $m_{S_\pm}$, $m_{\widetilde{R}_{2}} \equiv m_{\widetilde{R}_{2}^{(-1/3)}}$, and $m_{R_2} \equiv m_{R_2^{(5/3)}}$, where we define the mass eigenstates for mixed states $\widetilde{R}_2^{(2/3)}$ and $R^{(2/3)}_2$ as $S_{\pm} \equiv S_{\pm}^{(2/3)}$. 

We proceed by switching on only those Yukawa couplings in Eq.~\eqref{eq:R2tR2} that are absolutely necessary to generate contribution towards $(g-2)_\mu$ via the bottom quark exchange. These are $y_L^{b\mu}$ and $y_R^{b\mu}$, where we set all other entries in $y_L$ and $y_R$ matrices, as defined in Eqs.~\eqref{eq:sss} and~\eqref{eq:ttt}, to zero. The five relevant parameters for our numerical analysis are $m_{R_2}$, $m_{\widetilde{R}_2}$, $\xi$, $y_L^{b\mu}$, and $y_R^{b\mu}$. We show in the left panel of Fig.~\ref{fig:R2-final} the analysis of the maximal mixing scenario for $m_{R_2} =m_{\widetilde{R}_2}=1.6$\,TeV and $\xi=\sqrt{4 \pi}$ in the $y_L^{b\mu}$-$y_R^{b\mu}$ plane. The right panel of Fig.~\ref{fig:R2-final} shows the corresponding parameter space for $m_{R_2}=1.6$\,TeV, $m_{\widetilde{R}_2}=3$\,TeV, and $\xi=\sqrt{4 \pi}$.

It is clear from Fig.~\ref{fig:R2-final} that there is a tension within the $\widetilde{R}_2\,\&\, R_2$ scenario to simultaneously accommodate $(g-2)_\mu$ and all other relevant constraints even at the 2\,$\sigma$ level. Moreover, an improvement of the LHC limits from the study of $pp\to \mu\mu$ at high-$p_T$ would exclude all currently available parameter space.

\section{Conclusions}
\label{sec:conclusion}

We investigate viability of those scenarios where the one-loop contributions towards the anomalous magnetic moment of muon are induced through the mixing of two scalar LQs of the same electric charge $Q$ via the interactions with the SM Higgs field. The LQ pairs in question need to couple to muons and quarks of opposite chiralities in order to produce chirality-enhanced contributions. The three LQ pairs that satisfy these criteria are $S_1\,\&\, S_3$, $\widetilde{S}_1\,\&\, S_3$, and $\widetilde{R}_2\,\&\, R_2$, where the two states that mix have the electric charges $Q=1/3$, $Q=4/3$, and $Q=2/3$, respectively. 

We considered, in all three instances, the most minimal set of Yukawa couplings that can generate chirality-enhanced contribution to $(g-2)_\mu$ at the one-loop level. This contribution is proportional to the dimensionless mixing parameter and the product of one left- and one right-chiral coupling of LQs to the muons whereas it is inversely proportional to the LQ masses that we limit from below at $1.6$\,TeV for our numerical analysis.

In the $S_1\,\&\, S_3$ scenario the quarks in the loop are the up-type ones and we investigate viability of both the top quark and the charm quark cases, where, in addition to the electroweak precision measurement constraints, we impose relevant input from flavor physics, in particular $Z\to ll$, and the current LHC data analyses. It turns out that the electroweak precision measurement constraints are only marginally relevant for the LQ masses and associated Yukawa couplings that are not in conflict with the existing LHC studies. We proceed to find that the top-quark loop contributions represent a viable option to address $(g-2)_\mu$ whereas the charm-quark loop contributions are in direct conflict with the flavor physics constraints. If we allow, in the top quark case, that the relevant Yukawa couplings can reach the perturbativity limit, we find that the common LQ mass scale should be at or smaller than $5$\,TeV if one is to simultaneously address $(g-2)_\mu$ and satisfy other relevant constraints at the 1\,$\sigma$ level.

In the $\widetilde{S}_1\,\&\, S_3$ and $\widetilde{R}_2\,\&\, R_2$ scenarios the quarks in the $(g-2)_\mu$ loops are the down-type ones and we accordingly consider the bottom quark contributions only. We find, after imposing all relevant flavor physics and LHC constraints, that both scenarios not only require the two Yukawa couplings to be $\mathcal{O}(1)$ parameters to be able to address the $(g-2)_\mu$ discrepancy but exhibit tension with the existing LHC limits from the study of $pp\to \mu\mu$ at high-$p_T$. The projected limits of that study that correspond to the 300\,fb$^{-1}$ worth of the LHC data would significantly impair the ability of $\widetilde{S}_1\,\&\, S_3$ and $\widetilde{R}_2\,\&\, R_2$ scenarios to simultaneously accommodate the $(g-2)_\mu$ discrepancy and the most relevant flavor physics constraints.

\section*{Acknowledgments}
\label{sec:acknowledgments}

We thank D.\ Be\v{c}irevi\'{c}, N.\ Ko\v{s}nik, and P.~Paradisi for many useful discussions. I.D.\ acknowledges support of COST Actions CA15108 and CA16201. S.F.~acknowledges support of the Slovenian Research Agency through research core funding No.~P1-0035. This project has received support by the European Union's Horizon 2020 research and innovation programme under the Marie Sklodowska-Curie grant agreement N$^\circ$~674896 (ITN Elusives).

\end{document}